# Initial stages of the graphite-SiC(0001) interface formation studied by photoelectron spectroscopy

K.V. Emtsev[1,a], Th. Seyller[1,b], F. Speck[1], L. Ley[1],
P. Stojanov[2], J.D. Riley[2], R.G.C. Leckey[2]

[1] Lehrstuhl für Technische Physik, Universität Erlangen-Nürnberg, Erwin-Rommel-Strasse 1, D-91058 Erlangen, Germany

[2] Department of Physics, La Trobe University, Bundoora, Victoria 3083, Australia

[a] Konstantin.Emtsev@physik.uni-erlangen.de, [b] Thomas.Seyller@physik.uni-erlangen.de

**Abstract.** Graphitization of the 6H-SiC(0001) surface as a function of annealing temperature has been studied by ARPES, high resolution XPS, and LEED. For the initial stage of graphitization – the 6√3 reconstructed surface – we observe σ-bands characteristic of graphitic $sp^2$-bonded carbon. The π-bands are modified by the interaction with the substrate. C1s core level spectra indicate that this layer consists of two inequivalent types of carbon atoms. The next layer of graphite (graphene) formed on top of the 6√3 surface at $T_A$=1250°C-1300°C has an unperturbed electronic structure. The annealing at higher temperatures results in the formation of a multilayer graphite film. It is shown that the atomic arrangement of the interface between graphite and the SiC(0001) surface is practically identical to that of the 6√3 reconstructed layer.

**Introduction.**

The graphitization of the SiC surfaces by high temperature annealing in vacuum is known for a long time [1]. Monocrystalline graphite multilayers grow by the sublimation of Si from the SiC(0001) surface at $T_A$>1150°C. Several groups investigated the graphite-6H-SiC(0001) interface formation by surface science techniques such as X-ray photoelectron spectroscopy (XPS) [2], inverse photoelectron spectroscopy (IPES) [3], and scanning tunneling microscopy (STM) [4,5]. As an initial stage of graphitization a complex (6√3×6√3)R30° (6√3 hereafter) surface reconstruction occurs after annealing of 6H-SiC(0001) at $T_A$=1150°C. The nature of this reconstruction is discussed controversially. Fingerprints of graphitization (π* and σ* bands) were observed in IPES [3], while the energies of C1s surface shifted components differ from that of graphite [2]. The early models suggested a monolayer of graphite on top of either the unreconstructed surface or the (√3×√3)R30° reconstructed surface [1,3]. The latest model proposes that the surface is covered by graphene-like flakes arranged in a honeycomb structure [5]. In this work we have investigated the occupied electronic states of the 6√3 interfacial structure and ultra-thin graphite layers (few monolayers thick) formed on the SiC(0001) surface by angle-resolved photoelectron spectroscopy (ARPES), XPS, and lo- energy electron diffraction (LEED).

**Experiment.**

The samples used in this study were polished, on-axis cut 6H-SiC(0001) substrates with an n-type doping in the $10^{18}\,cm^{-3}$ range purchased from SiCrystal AG. The preparation of the graphitic layers on the SiC(0001) surface proceeds by vacuum annealing at elevated temperatures and results in the formation of a monocrystalline graphite film. Initially the Si-rich (√3×√3)R30° surface reconstruction was formed by annealing the sample in a flux of Si at $T_A$=1050°C. Such a surface is well characterized in the literature [6] (see also below). The transition from the Si- to the C-rich surface composition due to sublimation of the surface Si atoms takes place in the temperature range from 1050°C to 1150°C. The next ordered surface phase - 6√3 reconstruction – was obtained after annealing at $T_A$=1150°C for 5 min in vacuum. The subsequent annealing at $T_A$>1150°C was carried out in 50°C steps for Δt=2 min. After each annealing step the surfaces were characterized by LEED, ARPES, and XPS at the synchrotron light source BESSY II. Full ARPES measurements of the valence band states were performed at the beamlines TGM4 and UE56-2 using a Toroidal Electron Analyzer [7]. High resolution core level studies were performed by means of the Specs-PHOIBOS150 electron analyzer at the undulator beamline U49/2.

**Results and discussion.**

Fig.1 presents the photoemission intensity maps of the valence band states as a function of binding energy and electron momentum for the initial, Si-rich √3 reconstructed surface of 6H-SiC (a) and for the surfaces obtained during vacuum annealing with different graphite coverage (b-d). The coverage is determined predominantly by the annealing temperature.

For the initial √3 reconstructed surface of 6H-SiC(0001) (Fig.1(a)) we find characteristic surface state bands ($D_{Si}$ and $P_{1,2}$) in good agreement with theoretical calculations [8, and references therein]. These states originate from dangling bonds ($D_{Si}$) and backbonds ($P_{1,2}$) of the Si adatoms (1/3 of a monolayer) on top of the unreconstructed SiC(0001) surface. The other transitions visible are attributed to the SiC bulk bands.

In the temperature range from 1050°C to 1150°C the new 6√3 reconstruction starts to develop. As judged from LEED two phases (√3 and 6√3) coexist. This is also observed in the ARPES spectra (not shown). The transition is completed after annealing at $T_A$=1150°C for about 5 min. The LEED spots associated with the √3 reconstruction disappear at this temperature. The spectra taken from the 6√3 reconstructed surface are shown in Fig.1b. The graphitic nature of the 6√3 reconstruction becomes immediately apparent. We observe bands characteristic of the in-plane σ-bands of the $sp^2$-graphitic carbon. (In fact, σ-bands can be observed already after annealing at $T_A$=1100°C for the mixed (√3 and 6√3) surface, in agreement with the earlier IPES results [3]). SiC bulk bands are no longer visible. The σ-bands match the calculated bands of bulk graphite (grey line in Fig.1(b), from [9]) quite well but appear to be shifted by 1.0 eV to higher binding energies (bottom of the σ-band at 22.7 eV). From Fig. 1b it is clear that the graphitic layer of the 6√3 reconstruction is epitaxially ordered with the unit lattice rotated by 30° with respect to the SiC substrate. The same relation is observed for thicker graphite films. Interestingly, no graphitic π-bands are developed (compare to Fig. 1d). Instead, a few states with energy gaps are observed in the energy range where the π-bands are expected. This effect is most likely due to the overlayer-substrate interaction. The π-bands are formed by delocalized $p_z$-electrons of graphite and therefore are much more sensitive to any modification of the layer than the strong in-plane σ-bands. Also, no states can be detected at the Fermi level for the 6√3 surface. This is in contrast to the semi-metallic nature of graphite.

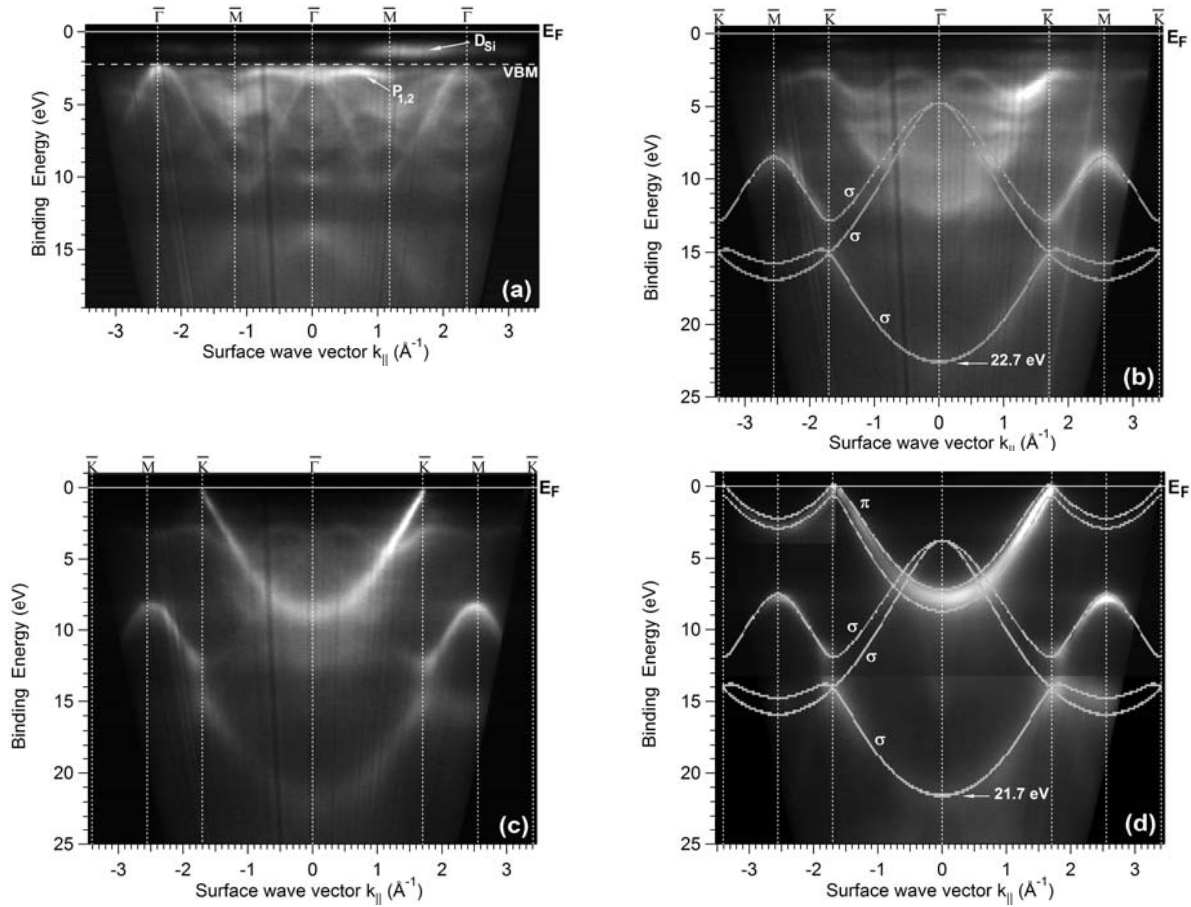

Fig.1 ARPES spectra taken on the initial 6H-SiC(0001)-√3 surface (a) and after several annealing steps corresponding to increasing graphite coverage: (b) 6√3 reconstructed surface, $T_A$=1150°C; (c) graphene layer on top of the 6√3 surface, $T_A$=1300°C; (d) bulk-like graphite $T_A$>1450°C. (Solid lines are the calculated bands for bulk graphite from Ref. [9]).

The single π-band emerges only after annealing at $T_A$=1250°C-1300°C (Fig. 1(c)). This indicates that the electronic structure of the graphite layer situated on top of the 6√3 surface is unperturbed and characterized by delocalized $p_z$-electrons. At this stage, states at the Fermi level are visible at the Brillouin zone boundary (<u>K</u>-point). These states arise from the slightly occupied π*-band.

Upon annealing at higher temperatures the growth of further graphite layers is manifested by the appearance of the second set of the π-bands due to the interlayer interaction in Bernal graphite formed with ABAB stacking (not shown). Finally, prolonged annealing at $T_A$>1450°C results in the formation of a ≈2 nm thick film (≈6 graphene layers) as estimated from the XPS data [10]. Such a film has an electronic structure indistinguishable from that of natural graphite (Fig. 1(d)).

The C1s spectrum of the 6√3 reconstructed surface is shown in Fig. 2 along with the result of its deconvolution. A minimum of three components are required to fit the data adequately. The component marked *SiC* at 283.70±0.10 eV binding energy is due to the SiC substrate signal. The other two components *S1* at 284.75±0.10 eV and *S2* at 285.55±0.10 eV (respective chemical shifts of 1.05 and 1.85 eV relative to the SiC bulk) are related to the carbon atoms located at the surface. No significant change in the intensity ratio (*S2/S1*) was observed with varying surface sensitivity of the experiment. Hence, the 6√3 overlayer consists of two inequivalent types of C atoms. The intensity ratio of the surface to bulk components ((*S1+S2*)/*SiC*) in Fig. 2 is relatively high. One fact, however, should be taken into account when estimating the thickness of the overlayer from this ratio. As is evident from the above ARPES data the 6√3 surface consists of *sp²*-bonded graphitic carbon and thus has a higher surface atomic concentration than SiC. Actually, the surface atomic concentration of C atoms in a layer of graphite is about 3 times that of SiC ($3.82 \cdot 10^{15}$ cm$^{-2}$ *vs.* $1.22 \cdot 10^{15}$ cm$^{-2}$). Applying a simple layer attenuation model we obtain the overlayer thickness of 2.4±0.3 Å. This value is smaller than the interlayer distance in graphite (3.35 Å) and closer to the bilayer separation in SiC (2.52 Å). Thus it appears that the surface coverage of the 6√3 reconstruction doesn't exceed a monolayer of graphitic carbon with an interfacial spacing shorter than in bulk graphite.

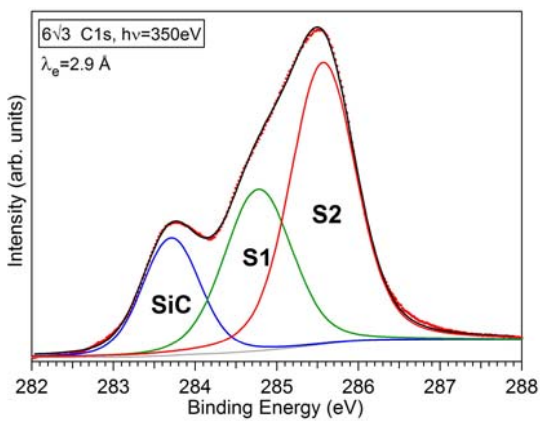

Fig. 2 The C1s core level spectrum and the result of its deconvolution for the 6√3 reconstructed surface.

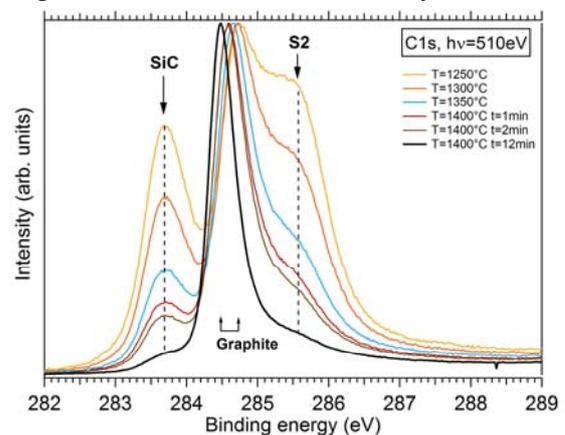

Fig. 3. The evolution of the C1s core level spectra upon graphitization. All spectra are normalized to the same maximum.

The evolution of the C1s spectra upon graphitization taken at a photon energy of 510 eV is shown in Fig. 3. As the annealing temperature increases the components *SiC* and *S2* are attenuated by the growing graphite layer. Finally, the last annealing step (bold line in Fig. 3) results in a formation of approximately 5ML-thick (≈1.7 nm) graphite film on top of the substrate. It is interesting to notice that during graphitization the ratio *S2/SiC* remains fairly constant. This implies that an atomic arrangement identical to the 6√3 reconstructed surface remains at the interface between graphite and SiC(0001). In other words, the 6√3 reconstruction is buried by the growing graphite without a major rearrangement of the atoms at the interface.

**Summary**

In conclusion, the growth of graphite on the 6H-SiC(0001) surface begins with the formation of the 6√3 reconstruction at $T_A$=1150°C for which the σ-bands characteristic of graphitic sp²-bonded carbon are detected. In contrast to that, the π-band is not developed. C1s core level spectra indicate that this layer consists of two inequivalent types of carbon atoms. Further annealing reveals the growth of a single monocrystalline graphite layer (graphene) on top of the 6√3 surface at T=1250°C-1300°C and multiple layer graphite at higher temperatures. The atomic arrangement of the graphite-SiC(0001) interface is practically identical to that of the 6√3 reconstruction.

**Acknowlegements**

We thank the staff of BESSY as well as G. Gavrila (TU Chemnitz) for support during experiments.